\documentclass[epj]{svjour}
\usepackage{graphics}
\begin{document}
\title{Mott Effect and $J/\psi$ Dissociation
at the Quark-Hadron Phase Transition}
\author{D. Blaschke\inst{1,2} \and 
G. Burau\inst{2} \and
Yu. Kalinovsky\inst{3}\thanks{\emph{Support: DFG grant No. 436 RUS 17/129/02}} 
\and 
T. Barnes\inst{4,5}\thanks{\emph{Support: NSF grant No. INT-0004089}}
}                     
%
%
\institute{ 
Department of Physics University of Rostock,
D-18051 Rostock, Germany
\and 
Bogoliubov Laboratory for Theoretical
Physics, JINR Dubna, 141980 Dubna, Russia
\and
Laboratory for Information Technologies, JINR Dubna,
141980 Dubna, Russia
\and 
Oak Ridge National Laboratory, Oak Ridge,
TN 37831-6373, USA
\and
Department of Physics and Astronomy, University of Tennessee,
Knoxville, TN 37996-1501, USA
}
\date{Received: date / Revised version: date}
%
\abstract{
We investigate the in-medium modification of pseudoscalar and
vector mesons in a QCD motivated chiral quark model by solving
the Dyson-Schwinger equations for quarks and mesons at finite
temperature for a wide mass range of meson masses, from light ($\pi$,
$\rho$) to open-charm ($D$, $D^*$) states.
At the chiral / deconfinement phase transition,
the quark-antiquark bound states enter the continuum
of unbound states and become broad resonances
(hadronic Mott effect).
We calculate the in-medium cross sections for charmonium dissociation
due to collisions with light hadrons in a chiral Lagrangian
approach, and show that the $D$ and $D^*$ meson spectral broadening lowers
the threshold for charmonium dissociation by
$\pi$ and $\rho$ mesons.
This leads to a step-like enhancement in the reaction rate.
We suggest that this mechanism for enhanced charmonium dissociation
may be the physical mechanism underlying
the anomalous $J/\psi$ suppression observed by NA50.
} 
\maketitle
\section{Introduction}
\label{intro}
Charmonia, in particular the $J/\psi$ meson, play an important role in the
experimental search for the quark-gluon plasma (QGP) in relativistic
heavy-ion collisions. The anomalous suppression of $J/\psi$ production
found by the NA50 collaboration at CERN SPS
in 158A~GeV Pb-Pb collisions
\cite{na50} is reminiscent of the signal for QGP formation
suggested by Matsui and Satz \cite{ms86}, although one should also
consider the competing, non-QGP mechanisms for $J/\psi$ suppression
such as charmonium dissociation through collisions with
projectile and target nucleons \cite{gh88} and
by comoving hadrons formed in the collision \cite{v88}.
A combination of hadronic and quark/gluon processes appears to give a
satisfactory parametrization of the NA50 data
\cite{Grandchamp:2002wp,Xu:2002zv}.
However, we should inquire as to whether
we do now have a consistent picture of anomalous $J/\psi$ suppression.
In this contribution we consider a unified theoretical
approach based on the quark (gluon) substructure of hadrons,
which predicts a characteristic energy
dependence of the $J/\psi$ dissociation cross sections
through collisions with light hadrons, as well as the
dissociative ``Mott effect" at the
chiral / deconfinement phase transition.
\section{The Mott effect and spectral function for D-Mesons}
\label{sec:1}
Due to their strong couplings to two-body decay channels,
light mesons such as the $\rho$
and the controversial light $\sigma$
can be modeled as quark-antiquark bound states or alternatively
as meson-meson interactions in the corresponding channel.
In the I=0 $\pi\pi$ ``$sigma$" channel,
the total spectral width $\Gamma_{\sigma}(T)$ associated with
a $\sigma$-meson shows a minimum that correlates
with the chiral restoration phase transition in
the phase diagram of strongly interacting matter \cite{kv+98}, since the
hadronic decay width $\Gamma_{\sigma \to 2 \pi}$ is already negligible but
the coupling $\Gamma_{\sigma \to q \bar q}$ is still small.
The transition from a bound state with vanishing decay width (infinite
lifetime) to a resonance in the continuum of unbound states is called
the Mott transition \cite{rr}, and can be described by the behavior of the
spectral function
\begin{equation}
\label{a}
A_h(s;T)=\frac{1}{N}\frac{\Gamma_h(T)~M_h(T)}
{[s-M_h^2(T)]^2+\Gamma^2_h(T)~M^2_h(T)}~,
\end{equation}
where
$M_h(T)$
and
$\Gamma_h(T)$
are the temperature-dependent mass and width
of the hadron $h$.
Critical phenomena related to the Mott transition for mesons at the chiral
transition have been
discussed in the context of the
NJL model for quark matter in Ref.\cite{hkr96}.
In this model it was found that the Mott transition temperature
for $D$ mesons is very close to that of the $\pi$ and $K$ mesons
\cite{gk92}. To estimate spectral function parameters
(\ref{a}) for the light and open-charm mesons, we use a modified NJL model
in which unphysical quark production thresholds below the Mott temperature
are excluded by an infrared cutoff \cite{bbvy}.
In the following section we investigate
the consequences of the meson Mott effect for charmonium dissociation
processes.
\section{In-medium $J/\psi$ dissociation cross section}
\label{sec:2}

\begin{figure}
\resizebox{0.4\textwidth}{!}{%
  \includegraphics{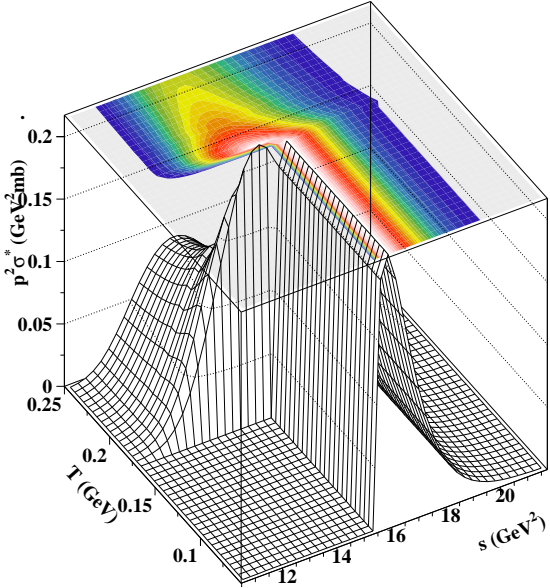} 
}
\caption{In-medium cross section $\sigma_h^*$ for
$J/\psi + \pi \to D^* + \bar{D}$
as function of $s$
and $T$, using
the spectral function of Ref.\cite{bbvy}.}
\label{sigmaT1}       
\end{figure}

\begin{figure}
\resizebox{0.4\textwidth}{!}{%
  \includegraphics{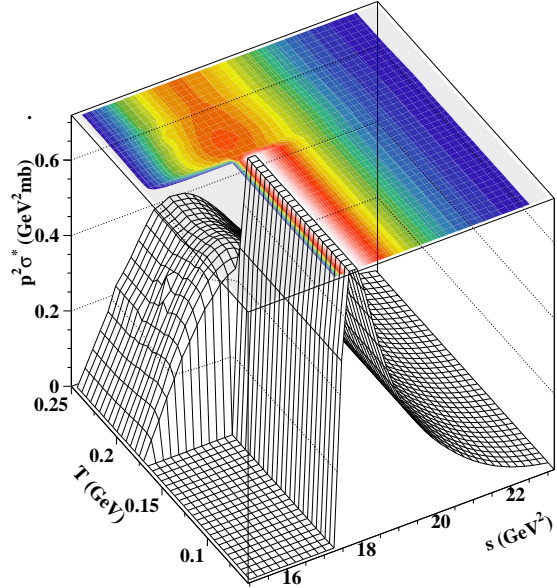}
}
\caption{Same as Fig. \ref{sigmaT1} for the process
$J/\psi + \rho \to D^* + \bar{D}^*$.}
\label{sigmaT2}       
\end{figure}

The in-medium dissociation
cross section is defined in the Green function formalism by
\begin{equation}
\sigma_{\psi h}^*(s;T)\!=\!\!\!\int\!\!ds_1\!\!\int\!\!ds_2~
A_{D_1}(s_1;T)~A_{D_2}(s_2;T)~\sigma_{\psi h}^{\rm vac}(s;s_1,s_2) .
\end{equation}
(For details of this approach see Ref.\cite{bbk}.)
The cross sections for the processes
$J/\psi + \pi \to D^* + \bar{D}$ and
$J/\psi + \rho \to D^* + \bar{D}^*$
are displayed in Figs. \ref{sigmaT1} and \ref{sigmaT2}.
The vacuum cross sections
$\sigma_{\psi h}^{\rm vac}$ assumed here follow
from the chiral Lagrangian approach of Ref.\cite{ikbb}.

\begin{figure}[h]
\resizebox{0.4\textwidth}{!}{%
  \includegraphics{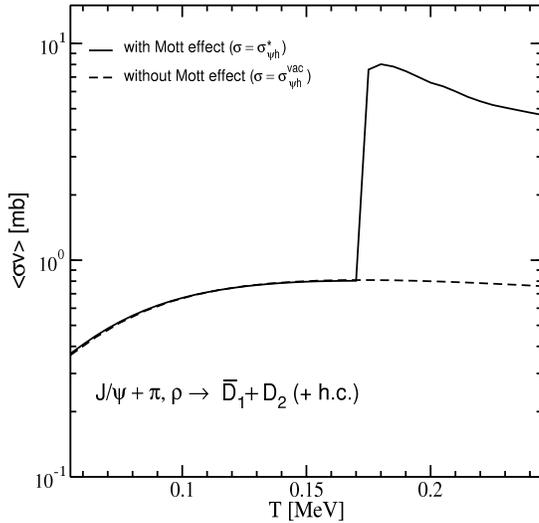}
}
\caption{Total thermal averaged $J/\psi$ dissociation cross section 
as function of $T$ without (dashed line) and with Mott effect for the 
$\pi$-, $\rho$ and $D$-mesons (solid line), using the spectral functions 
of Ref.\cite{bbvy}.}
\label{sigvT}       
\end{figure}

In both cases the Mott effect
($T^{\rm Mott} \approx 172~{\rm MeV}$) increases the spectral width in
(\ref{a}) of the $D$ and $D^*$ mesons, and hence effectively lowers the
threshold for charmonium dissociation reactions. At temperatures above
$T^{\rm Mott}$ these reactions become exothermic, whereas they were
endothermic for $T < T^{\rm Mott}$.
In addition to the lowered reaction thresholds, in Figs. \ref{sigmaT1}
and \ref{sigmaT2} one can also
see a decrease in the maximum values of these exclusive cross sections
multiplied by the squared momentum $p^2$ of the incoming light particles. 
This behavior is slightly larger in
$J/\psi + \pi$
than
$J/\psi + \rho$
at high temperatures. In general, $J/\psi + \rho$ is the dominant
dissociation process.
In Ref.\cite{b3ks} it was already shown that this behavior of the in-medium
cross section leads to a strong enhancement in the thermal-averaged
dissociation cross section, {\it i.e.}
in the inverse lifetime of the $J/\psi$ given by 
$\tau^{-1} = \tau^{-1}_{\pi} + \tau^{-1}_{\rho}$, with
\begin{eqnarray}
\tau^{-1}_{h} &=& \langle \sigma_{\psi h}^* v \rangle~n_{h}(T)\\
&=&\! \int\! \frac{d^3p}{(2\pi)^3}\! \int ds'\!
A_{h}(s';T)~f_{h}(p,s';T)\! j_{h}(p,s')\!\sigma_{\psi h}^*(s;T)~,\nonumber
\end{eqnarray}
where $f_{h}(p,s';T)$ is the Bose distribution function with the
energy argument $E(p,s') = [p^2 + s']^{1/2}~$, and $j_{h}(p,s')$ is the
flux factor for the $\psi$-$h$ collisions, $h = \{ \pi, \rho\}$. 
The thermal average of this total $J/\psi$ dissociation cross section 
as function of $T$, using the in-medium cross sections shown in 
Figs. \ref{sigmaT1} and \ref{sigmaT2}, is displayed in Fig. \ref{sigvT}.
This $J/\psi$ dissociation rate due to impact by hadronic resonances 
shows a step-like enhancement by an order of magnitude above 
$T^{\rm Mott}$ for mesonic states due to their spectral broadening and 
thus the effective lowering of the breakup threshold. 
The effect is dominated by the $\rho$ meson subprocess shown in Fig. 
\ref{sigmaT2} and its magnitude is quite 
sensitive to the detailed temperature dependence of the $D$- and 
$D^*$-meson spectral functions, as was discussed in \cite{b3ks}.

The hadronic Mott effect scenario described here
could be the dominant physical mechanism underlying
the anomalous $J/\psi$ suppression \cite{na50}.
It could also contribute to an understanding of fast chemical 
equilibration \cite{pbm} reported for the NA50 results on charm production. 
The role of the Mott effect on charmonium recombination 
($D \bar{D}$ fusion) is an interesting topic for further investigation.
It remains to be investigated which role dissociation processes by quark and 
gluon impact in the plasma phase have to play. 
A preliminary comparison with the NA50 data \cite{gbdiss} shows that the 
bulk of the the anomalous $J/\psi$ suppression pattern can be explained 
by dissociation due to hadronic resonance impact provided the dramatic changes
of mesonic spectral functions obtained from the modified NJL model calculation
\cite{bbvy} turn out to be realistic.
\section{Conclusion}%
An understanding of
the quark substructure of mesons is essential for determining
meson-meson vacuum cross sections as well as their modifications in
hadronic matter.
We have shown that the $D$-meson Mott effect
at the QGP phase transition reduces the threshold for charmonium
dissociation, which leads to a large increase in the
$J/\psi$ dissociation rate.
A new finding reported here
is that $J/\psi$ dissociation is dominated by $J/\psi + \rho$ collisions,
so that a direct connection to the in-medium $\rho$ spectral function, as
measured for example by NA45 (CERES), should be investigated.
In subsequent research we plan to model the temperature-dependent
meson spectral functions, using QCD Dyson-Schwinger
equations as an improvement over the NJL model.
Comparison with recent lattice QCD results for these spectral functions
\cite{Karsch:2002wv} is a promising approach for future studies of the
hadronic Mott effect at the quark-hadron phase transition.
\section*{Acknowledgements}
We are grateful to many colleagues for discussions, in particular to
P.-B. Gossiaux, J. H\"ufner, C.-M. Ko, S.H. Lee, Y. Oh, P. Petreczky,
A. Polleri, R. Rapp and C.Y. Wong.
G.B. was supported by the DFG Graduiertenkolleg
``Stark korrelierte Vielteilchensysteme'', and
D.B. and G.B. acknowledge support from the DAAD
for their visits to Oak Ridge National Laboratory
and the University of Tennessee at Knoxville.
%
%

\end{document}